\documentclass[smallextended]{svjour3}       
\smartqed  
\usepackage{graphicx}
\usepackage{psfrag}
\usepackage{longtable}
\usepackage{mathptmx}      
\journalname{Experimental Astronomy}
\newenvironment{itemlist}{
\begin{list}{}{\setlength{\leftmargin}{10mm}\setlength{\parsep}{0mm}
\setlength{\itemsep}{3mm}} }{\end{list}}

\begin{document}

\title{Herschel celestial calibration sources}
\subtitle{Four large main-belt asteroids as prime flux calibrators for the far-IR/sub-mm range}

\titlerunning{Asteroids as prime flux calibrators for Herschel}

\author{Thomas M\"uller      \and
        Zolt\'an Balog       \and
        Markus Nielbock      \and
        Tanya Lim            \and
	David Teyssier       \and
        Michael Olberg       \and
        Ulrich Klaas         \and
        Hendrik Linz         \and
        Bruno Altieri        \and
        Chris Pearson        \and
        George Bendo         \and
        Esa Vilenius
}


\institute{T. M\"uller, E. Vilenius \at
           Max Planck Institute for Extraterrestrial Physics, PO Box 1312, Giessenbachstrasse,
           85741 Garching, Germany \\
           Tel.: +49 89 30000-3499\\
           \email{tmueller@mpe.mpg.de}
           \and
           Z. Balog, M. Nielbock, U. Klaas, H. Linz \at
           Max Planck Institute for Astronomy, K\"onigstuhl 17, 69117 Heidelberg, Germany
           \and
           T. Lim, C. Pearson \at
           Space Science and Technology Department, RAL, Didcot, OX11 0QX, Oxon UK
           \and
           D. Teyssier, B. Altieri \at
           European Space Astronomy Centre (ESAC), ESA, Villanueva de la Ca\~nada, 28691 Madrid, Spain
           \and
           M. Olberg \at
           Onsala Space Observatory, Chalmers University of Technology, 43992 Onsala, Sweden   
           \and
           G. Bendo \at
           UK ALMA Regional Centre Node, Jodrell Bank Centre for Astrophysics, Manchester M13 9PL, United Kingdom
	   \\
}

\date{Received: date / Accepted: date}

\maketitle

\begin{abstract}
Celestial standards play a major role in observational astrophysics. They are needed to characterise the
performance of instruments and are paramount for photometric calibration. During the
Herschel Calibration Asteroid Preparatory Programme approximately 50 asteroids have
been established as far-IR/sub-mm/mm calibrators for Herschel. The selected asteroids
fill the flux gap between the sub-mm/mm calibrators Mars, Uranus and Neptune, and the
mid-IR bright calibration stars. All three Herschel instruments observed asteroids for various
calibration purposes, including pointing tests, absolute flux calibration, relative
spectral response function, observing mode validation, and cross-calibration aspects.
Here we present newly established models for the four large and well characterized
main-belt asteroids (1)~Ceres, (2)~Pallas, (4)~Vesta, and (21)~Lutetia which
can be considered as new prime flux calibrators.
The relevant object-specific properties (size, shape, spin-properties, albedo,
thermal properties) are well established. The seasonal (distance to Sun, distance to observer,
phase angle, aspect angle) and daily variations (rotation) are included in a new thermophysical
model setup for these targets. The thermophysical model predictions agree within
5\% with the available (and independently calibrated) Herschel measurements.
The four objects cover the flux regime from just below 1,000\,Jy (Ceres at mid-IR N-/Q-band)
down to fluxes below 0.1\,Jy (Lutetia at the longest wavelengths). Based on the comparison with
PACS, SPIRE and HIFI measurements and pre-Herschel experience, the validity
of these new prime calibrators ranges from mid-infrared to about 700\,$\mu$m, connecting nicely
the absolute stellar reference system in the mid-IR with the planet-based calibration
at sub-mm/mm wavelengths.
\keywords{
	Herschel Space Observatory \and
	PACS \and
	SPIRE \and
	HIFI \and
	Far-infrared \and 
	Instrumentation \and 
	Calibration \and
	Celestial standards \and
        Asteroids
}
\end{abstract}

\section{Introduction}
\label{intro}

\begin{figure}[h!tb]
\centering
  \rotatebox{90}{\resizebox{!}{\hsize}{\includegraphics{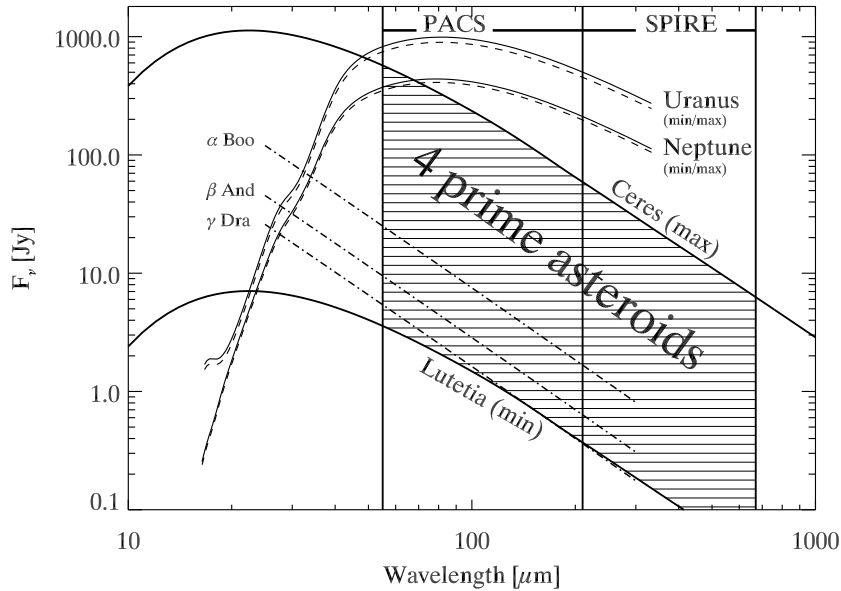}}}
  \caption{Overview with the flux densities of the different far-IR/sub-mm/mm calibrators.
           The Uranus and Neptune SEDs represent the minimum and maximum fluxes during
           Herschel visibility phases. Three fiducial stars are also shown, their flux
           coverage is representative for the brightest stellar calibrators.
           For Ceres and Lutetia we show the minimum and maximum fluxes during Herschel
           observations.}
\label{fig:starsasteroidsplanets}
\end{figure}

With the availability of the full thermal infrared (IR) wavelength range (from a
few microns to the millimetre range) through balloon, airborne and
spaceborne instruments, it became necessary to establish new calibration
standards and to develop new calibration strategies. Instruments working
at mm-/cm-wavelengths were mainly calibrated against the planets
Mars, Uranus and Neptune (e.g., \cite{griffin93,moreno98,orton03,sandell03,griffin13}),
while the mid-IR range was always tied to stellar models (e.g.,
\cite{rieke85,hammersley03,cohen03,decin03,engelbracht10,gordon10,reach10,wright10,dehaes11}).
For the far-IR/sub-mm regime no optimal calibrators were available right from
the beginning: the stars are often too faint for calibration aspects which require high
signal-to-noise (S/N) ratios or are problematic in case of near-IR filter
leaks\footnote{Near-IR filter leaks are photometrically problematic when near-IR bright
objects -like stars- are observed in far-IR bands.}.
The planets are very bright and not point-like anymore. They are causing
saturation or detector non-linearity effects. Between these two
types of calibrators there remained a gap of more than two orders of
magnitude in flux. This gap was filled by sets of well-known and well-characterized
asteroids and their corresponding model predictions
(e.g., \cite{beichman88,mueller02,mueller03,stansberry07,kawada07,mueller05a}).
Figure~\ref{fig:starsasteroidsplanets} shows the flux-wavelength regime
covered by the three types of objects typically used for calibration
purposes at far-IR/sub-mm/mm wavelength range.

The idea of using asteroids for calibration purposes goes back to IRAS
\cite{beichman88}.
The IRAS 12, 25 and 60\,$\mathrm{\mu m}$ bands were calibrated via stellar
models and in that way connected to groundbased N- and Q-band measurements.
But at 100\,$\mu$m neither stellar model extrapolations nor planet models
were considered reliable. Asteroids solved the problem. Models for a selected
sample of large main-belt asteroids were used to ``transfer'' the observed
IRAS 60\,$\mu$m fluxes out to 100\,$\mathrm{\mu m}$ and calibrate in that way
the IRAS 100\,$\mu$m band \cite{beichman88}.

There was an independent attempt to establish a set of
secondary calibrators at sub-millimetre (sub-mm) wavelengths to fill
the gap between stars and planets \cite{sandell94}. Ultra-Compact H-II regions,
protostars, protoplanetary nebulae and AGB-stars were selected, but often these sources
are embedded in dust clouds which produce a strong and sometimes variable
background. The modeling proves to be difficult and accurate far-IR extrapolations
are almost impossible.

The Infrared Space Observatory (ISO) \cite{kessler96}
was also lacking reliable photometric standards at far-IR wavelength
(50 - 250\,$\mu$m) in the flux regime between the stars 
\cite{hammersley03,cohen03,decin03}
and the planetary calibrators Uranus and Neptune \cite{griffin93,orton03,lim03,klaas03,schulz03}.
M\"uller \& Lagerros \cite{mueller98} provided a set of 10 asteroids,
based on a previously developed thermophysical model code by Lagerros
\cite{lagerros96,lagerros97,lagerros98}. These sources have been 
extensively observed by ISO for the far-IR photometric calibration,
for testing relative spectral response functions and for many technical
instrument and satellite purposes.

AKARI \cite{murakami07} followed the same route to calibrate the Far-Infrared
Surveyor (FIS) \cite{kawada07} via stars, asteroids and planets in the
wavelengths regime 50 - 200\,$\mu$m.

The Spitzer mission \cite{werner04} considered in the beginning only stars
for calibration purposes. But due to a near-IR filter leak of the MIPS \cite{rieke04}
160\,$\mu$m band, the calibration scientists were forced to establish and verify
calibration aspects by using cooler objects. The asteroids served as reference
for the flux calibration of the 160\,$\mu$m band as well as for testing the
non-linear MIPS detector behaviour \cite{stansberry07}.

In preparation for Herschel \cite{pilbratt10} and
ALMA\footnote{\tt http://en.wikipedia.org/wiki/Atacama\_Large\_Millimeter\_Array}
a dedicated asteroid programme was established \cite{mueller05a}. This
led to a sample of about 50 asteroids for various calibration purposes.
Along the mission only the 12 asteroids with the highest quality characterization
were continued to be observed for calibration.

Here we present the Herschel observations and the data reduction of all photometrically
relevant asteroid measurements (Section~\ref{sec:obs}). First, the asteroid instrumental fluxes
in engineering units were converted to absolute fluxes using conversion factors derived from
stellar calibrators (PACS), Neptune (SPIRE) and Mars (HIFI). Next, the absolute fluxes were
corrected for differences in spectral energy distribution between the prime calibrator(s)
and the asteroids to obtain mono-chromatic flux densities at predefined reference wavelengths.
In Section~\ref{sec:tpm} we document recently updated asteroid models for Ceres, Pallas, Vesta,
and Lutetia. The models are entirely based on physical and thermal properties
taken from literature and derived from independent measurements, like occultation
measurements, HST\footnote{Hubble Space Telescope}, adaptive optics, or flyby missions.
We compare (Section~\ref{sec:res})
the absolute model predictions with all available photometric Herschel (PACS \cite{poglitsch10},
SPIRE \cite{griffin10}, HIFI \cite{degraauw10}) measurements
and discuss the validity and limitations. The dispersion in the ratio of model to measured
fluxes for the four asteroids determines the error in the calibration factor.
The conclusions are given in Section~\ref{sec:con}.
It is important to note here that the
derived Herschel flux densities of the asteroids were independently calibrated against
5 fiducial stars (PACS), the planet Neptune (SPIRE) and the planet Mars (HIFI). The 
asteroids are therefore also serving as unique cross-calibration objects between
the different calibration concepts, the different instruments, observing modes,
wavelengths- and flux regimes.


\section{Observations \& data reduction}
\label{sec:obs}

\subsection{PACS photometer observations}
\label{sec:obspacs}

The Photodetector Array Camera and Spectrograph (PACS
\cite{poglitsch10}) on board the Herschel Space Observatory
\cite{pilbratt10} provides imaging and spectroscopy capabilities. Here,
we only considered photometric measurements of the four asteroids, taken
with the imaging bolometer arrays either at 70/160\,$\mu$m (blue/red)
or at 100/160\,$\mu$m (green/red). Most of the observations were taken
as part of calibration programmes, with the majority of the measurements
taken in high gain and only a few in low gain.
The observations have either been taken in scan-map mode or in chop-nod mode.
The data were reduced in a standard way, following the steps defined in
the officially recommended chop-nod and scan-map reduction scripts,
described in more detail in \cite{balog13,nielbock13},
with flagging of bad and saturated pixels. The calibration was
based on the latest versions of the bolometer response file
(responsivity: FM,7) and the flat-fielding (flatField: FM,3). The
non-linearity correction was needed, with correction of up to 6\% for
the highest asteroid fluxes. The bolometer signals were also corrected for the
evaporator temperature effect (see \cite{moor13}), with correction
factors of -0.3\% to 3.2\%. Since the asteroids have apparent
Herschel-centric motions of up to 80$^{\prime \prime}$/h, the frames were
projected in an asteroid-centered (co-moving) reference frame for
the final maps. We used map pixel sizes of 1.1$^{\prime \prime}$,
1.4$^{\prime \prime}$, and 2.1$^{\prime \prime}$ at 70, 100, and 160\,$\mu$m,
respectively, to sample the point spread function in an optimal way.

\paragraph{PACS bolometer scan-map observations.}

The scan-map observations of the four prime asteroids are listed in
Tables~\ref{tbl:ceresobs1}, \ref{tbl:pallasobs1}, \ref{tbl:vestaobs1}, \ref{tbl:lutetiaobs1}
(observing mode 'PACS-SM').
They were obtained using the mini scan-map mode \cite{pacsom,mueller11c} with
the telescope scanning at a speed of 20$^{\prime \prime}$/s along
parallel legs of about 3$^{\prime}$-4$^{\prime}$ length. Typically 10 scan-legs,
separated by a few arcseconds, have been taken. The scans were performed
in such a way that the source was moving along the array diagonals
(70$^{\circ}$ and 110$^{\circ}$ scan-angles in array coordinates)
for optimized coverage and sensitivity. The data were reduced in a standard way (see
above), with details about the masking, high-pass filtering, speed selection
and deglitching of the data given in \cite{balog13}. We constructed final
images in the asteroid co-moving reference frame for each individual
OBSID\footnote{Herschel unique observation identifier} and band.
The combination of the scan and cross-scan observations was
not necessary for our bright point-sources.

\paragraph{PACS bolometer chop-nod observations.}

The chop-nod observations of the four prime asteroids are listed in
Tables~\ref{tbl:ceresobs2}, \ref{tbl:pallasobs2}, \ref{tbl:vestaobs2}, \ref{tbl:lutetiaobs2}
(observing mode 'PACS-CN').
They were obtained using the point-source photometry Astronomical
Observing Template (AOT) that is carried out by chopping and nodding
in perpendicular directions and with amplitudes of 52$^{\prime \prime}$
(see \cite{pacsom,mueller11c,nielbock13} for further details).
The data reduction included -in addition to what has been done for
scan-maps- an adjustment for the apparent response drift and offset in
this mode (see \cite{nielbock13}) resulting in corrections
of 4.3\% to 7.6\% for the four asteroids, depending on the time of
the observation (before/after OD\,300) and the band \cite{nielbock13}.
We produced final point-source maps in the asteroid co-moving reference frame.

\paragraph{Aperture photometry.}

We applied aperture photometry with radii of 12$^{\prime \prime}$ in
blue/green and 22$^{\prime \prime}$ in red, centered on the source image
in the final maps. Depending on the band, 78-82\% of the
source flux is inside these apertures (see \cite{lutz12}). The sky
noise in scan-maps was determined in a sky annulus with inner and outer
radii of 35$^{\prime \prime}$ and 45$^{\prime \prime}$, respectively
(see \cite{balog13}).
The sky noise in the chop-nod images was taken from a sky annulus with
inner/outer radii of 20$^{\prime \prime}$/25$^{\prime \prime}$ in the
blue and green maps and 24$^{\prime \prime}$/28$^{\prime \prime}$ in
red maps (see \cite{nielbock13}).
We performed colour corrections \cite{mueller11b} of 1.00, 1.02\footnote{The Vesta SED requires a
colour correction value of 1.03 in the green band}, and 1.07 for the blue,
green, and red band data to obtain monochromatic flux densities at the
PACS key wavelength of 70.0, 100.0, and 160.0\,$\mu$m, respectively\footnote{The PACS
photometric calibration is based on the assumption of a constant energy spectrum
of the observed source $\nu \times F_{\nu} = \lambda \times F_{\lambda}$. Asteroid SEDs
deviate from this assumption and colour-corrections are required}.
These corrections were calculated on basis of model SEDs for the four
asteroids and correspond roughly to the corrections for a 200-300\,K black
body.
The absolute flux uncertainties were calculated by adding
quadratically the measured sky noise (corrected for correlated noise, see
\cite{balog13}), 1\% for the uncertainties in colour correction, and
5\% for errors related to the fiducial star models which are the
baseline for the absolute flux calibration of the PACS photometer.
The derived flux densities and errors (typically around 5-6\%) are given in
Tables~\ref{tbl:ceresres}, \ref{tbl:pallasres}, \ref{tbl:vestares}, \ref{tbl:lutetiares}
together with the observing log and conditions (observation mid-time,
object distance from Sun and Herschel, phase angle).

\subsection{SPIRE photometer observations}
\label{sec:obsspire}

The SPIRE \cite{griffin10,swinyard10} photometer observations of the four prime asteroids are listed in
Tables~\ref{tbl:ceresobs3}, \ref{tbl:pallasobs3}, \ref{tbl:vestaobs3}, \ref{tbl:lutetiaobs3}.
The data were taken in four different observing modes
"Sm Map" (small map), "Lg Map" (large map), "Scan" (scan map), "PS" (point-source),
mainly as part of calibration programmes,
but here we only use data taken in the standard small and large map modes.
The measurements were reduced through HIPE\footnote{HIPE is a joint development by the Herschel Science
Ground Segment Consortium, consisting of ESA, the NASA Herschel Science Center, and the HIFI, PACS
and SPIRE consortia.} version 11 -using the SPIRE Calibration Tree version 11- by SPIRE
instrument experts and calibrated against a reference Neptune model (ESA4)
\cite{moreno12,bendo13,pearson13}. The processing of different observing modes is
essentially identical, with the exception of the so-called cooler burp
correction which was only done for large maps in a dedicated interactive analysis step.
Point source photometry was extracted from Level~1 data using the
timeline fitter task \cite{bendo13}, fitting an elliptical Gaussian to the asteroid in
the co-moving reference frame.
The object fluxes have been colour-corrected assuming
a spectral index of 2 \cite{spireom} to obtain
mono-chromatic flux densities at 250, 350, and 500\,$\mu$m
(corrections are in the order of 5-6\% here).
More details about the reduction and calibration are given in
Section 6.1 in \cite{griffin13}. 
The absolute flux uncertainties were calculated by adding
quadratically the individual errors from the Gaussian timeline fitting
(typically well below 1\%), 2\% for the uncertainties in colour correction,
and 5\% for errors related to the Neptune model
which is the baseline for the absolute flux calibration of the SPIRE photometer.
The derived flux densities and errors (typically around 5-6\%) are given in
Tables~\ref{tbl:ceresres}, \ref{tbl:pallasres}, \ref{tbl:vestares}, \ref{tbl:lutetiares}
together with the observing log and conditions (observation mid-time,
object distance from Sun and Herschel, phase angle).

\subsection{HIFI continuum observations}
\label{sec:obshifi}

The HIFI \cite{degraauw10} point-source observations of Ceres are listed
in Table~\ref{tbl:ceresobs4}. The data were taken \cite{hifiom} in
band~1a (OD~1392) and band~1b (ODs 923, 1247, 1260) as part of two
science programmes.  Here we only consider the continuum fluxes of
Ceres, which are a by-product of the data reduction and which were
originally not considered as relevant for the science case.
The HIFI continuum fluxes were derived
from the observations taken during the four ODs. For each of the four
data sets (i.e.\ for the 10\,h of integration on OD~1392) all data from
both polarizations -H and V- have been averaged.  The conversion of
double-sideband antenna temperatures to flux densities uses values for
the aperture efficiencies derived from observations of Mars and are
therefore tied to a model of this planet\footnote{\tt
  http://www.lesia.obspm.fr/perso/emmanuel-lellouch/mars/}.  The
aperture efficiencies have recently been reviewed and we use the new
values kindly provided to us by Willem Jellema (priv.\ comm.), which, at
the frequencies considered here, are smaller by 3.8\% and 5.9\% in the H
and V polarization, respectively, than values quoted in
\cite{roelfsema12}.  The given flux densities are averages of both
polarizations (i.e.\ H and V) observed by HIFI and are derived from
the median values of the observed continuum baselines. The error
calculation takes into account the noise r.m.s.\ after smoothing to a
resolution of 100\,MHz, quadratically added to the estimated 5\% error
in the Mars model which we tie our calibration to.
For continuum measurements the side-band ratio errors are negligible, and
standing wave effects are also averaged out. The two polarizations of
HIFI in band~1 are misaligned by $6.6^{\prime\prime}$, leading to a
coupling loss of order 2\% (for a perfect Gaussian beam and no satellite
pointing error).
Allowing for additional pointing errors, we estimate that the derived
flux densities could be too low by $\approx 5$\%. The derived flux
densities and errors are given in Table~\ref{tbl:ceresres} together with
the observing log and conditions (observation mid-time, object distance
from Sun and Herschel, phase angle).

\section{Thermalphysical model and asteroid-specific model parameters}
\label{sec:tpm}

The applied thermophysical model (TPM) is based on the work by
Lagerros \cite{lagerros96,lagerros97,lagerros98}.
This model is frequently and successfully applied to near-Earth
asteroids (e.g., \cite{mueller04b,mueller05b,mueller11a,mueller12}),
to main-belt asteroids (e.g., \cite{mueller98,mueller04a,orourke12}),
and also to more distant objects (e.g., \cite{horner12,lim10}).
The TPM takes into account the true observing and illumination geometry
for each observational data point, a crucial aspect for the interpretation
of the main-belt asteroid observations which cover a wide range of phase angles
and helio-/observer-centric distances, as well as different spin-axis
obliquities.

High quality size and geometric albedo values are fundamental 
for reliable TPM predictions. For all four asteroids we used
literature values, but only after a critical inspection of the
published sizes and albedos and their error estimates. The TPM also
allows one to specify simple or complex shape models
and spin-vector properties. The one-dimensional vertical heat conduction
into the surface is controlled by the thermal inertia $\Gamma$\footnote{The thermal inertia $\Gamma$ is defined
as $\sqrt{\kappa \rho c}$, where $\kappa$ is the thermal conductivity, $\rho$ the
density, and $c$ the heat capacity.}. The observed mid-/far-IR/sub-mm
fluxes are connected to the hottest regions on the asteroid surface and
dominated by the diurnal heat wave. The seasonal heat wave is less important
and therefore not considered here. The infrared beaming effects (similar to
opposition effects at optical wavelengths) are calculated
via a surface roughness model, described by segments of hemispherical craters.
Here, mutual heating is included and the true crater illumination and
the visibility of shadows is considered.

For the calculation of the Bond albedo (which is assumed to be close to the
bolometric albedo) also the object specific slope parameters for the phase curve G
and the absolute magnitudes H are needed (IAU two-parameter magnitude system for
asteroids \cite{lebofsky86,bowell89}). The Bond albedo is given by p$_V\cdot$q,
with the geometric V-band albedo p$_V$, and the phase integral q = 0.290 + 0.684$\cdot$G.
In cases where p$_V$ was measured in-situ, only G is required, in cases where p$_V$ was
not directly measured, we derived p$_V$ from H$_V$ and the object's effective size D$_{eff}$
via: p$_V$ = 10$^{(2\cdot\,log_{10}(S_0) - 2\cdot\,log_{10}(D_{eff}) - 0.4\cdot\,H_V)}$,
with the Solar constant, S$_0$ = 1361\,W/m$^2$.
We used literature values for H-G based on large samples of
measurements covering many aspect and phase angles from several apparitions.
Typical uncertainties in G values have a negligible influence
on the TPM flux predictions over the entire Herschel wavelength-range and for
the accessible phase angles (approximatly 15-30$^{\circ}$) for main-belt asteroids.
Errors in H are directly influencing the geometric albedo: 0.05\,mag in
H translate into a 5\% error in albedo, with a corresponding flux change well below 1\%.

The level of roughness is driven by the r.m.s.\ of the surface slopes which
correspond to a given crater depth-to-radius value combined with the fraction
$f$ of the surface covered by craters, see also Lagerros (\cite{lagerros96}) for
further details. For all four targets we used the ``default" roughness settings
($\rho$=0.7, f=0.6) \cite{mueller99}.

We used wavelength-dependent emissivity models with emissivities of 0.9 up to
150\,$\mu$m and slowly decreasing values beyond $\sim$150\,$\mu$m. The ``default" model
-used for Ceres, Pallas, and Lutetia- has lowest emissivities of around 0.8 in
the sub-mm-range, the Vesta-specific emissivity model is more extreme and has
values going down to 0.6 at 600\,$\mu$m. Both models are used as specified
and applied in \cite{mueller98,mueller99,mueller02}.

For the thermal inertia $\Gamma$ we used a ``default" value for large, regolith-covered
main-belt asteroids, namely $\Gamma$ = 15\,Jm$^{-2}$s$^{-0.5}$K$^{-1}$ \cite{mueller99}.
This value is not very well constrained and in the literature one can find smaller
values down to 5\,Jm$^{-2}$s$^{-0.5}$K$^{-1}$ (e.g., \cite{orourke12}) or
larger values of 25\,Jm$^{-2}$s$^{-0.5}$K$^{-1}$ (e.g., \cite{mueller98}).
The precise value has very little influence on far-IR and sub-mm-fluxes \cite{mueller02a}
 -at least for large regolith-covered main-belt asteroids- and it agrees very well with the
lunar value of $\Gamma$ = 39\,J\,m$^{-2}$\,s$^{-0.5}$\,K$^{-1}$ \cite{keihm84} considering
the lower temperature environment at 2-3\,AU from the Sun which lowers the thermal
conductivity within the top surface dust layer considerably.


\clearpage
\subsection{(1)~Ceres}

\begin{figure}[h!tb]
\centering
  \rotatebox{0}{\resizebox{5.5cm}{!}{\includegraphics{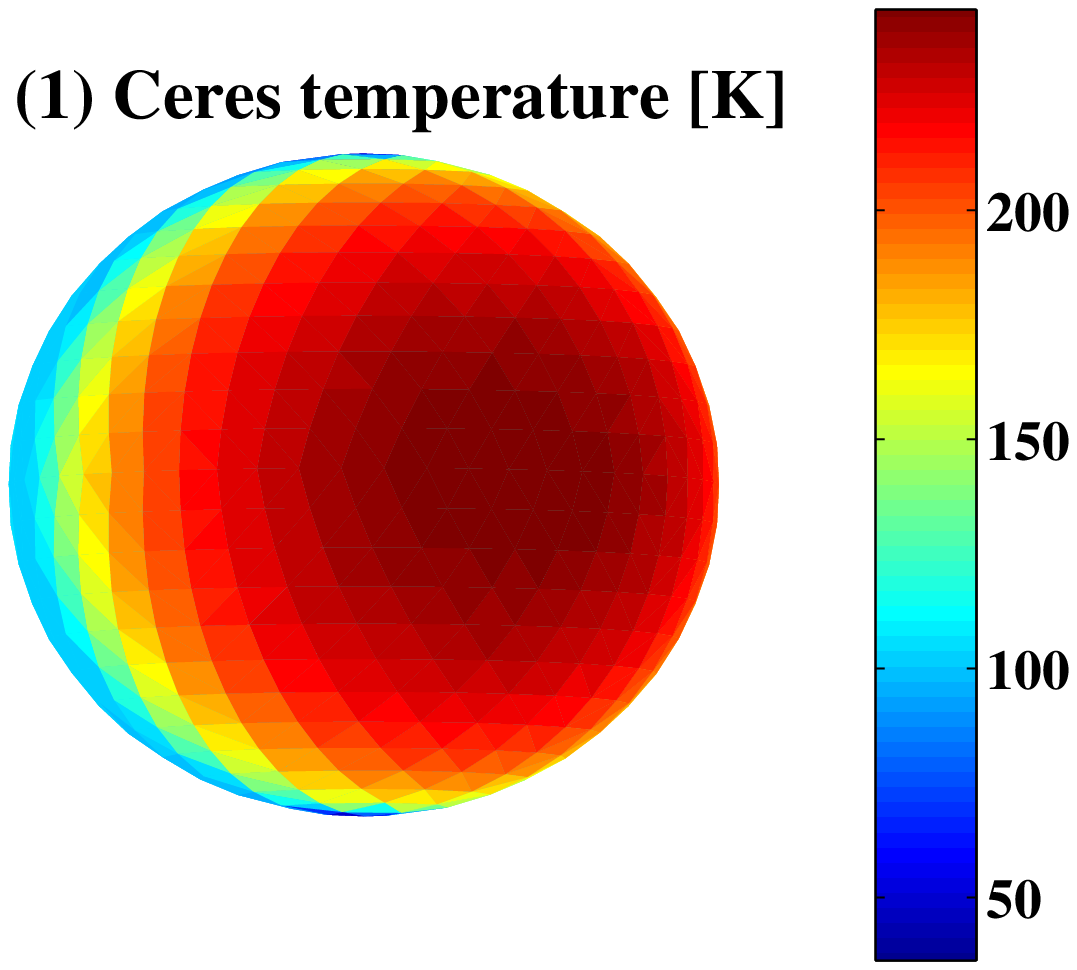}}}
  \rotatebox{0}{\resizebox{5.5cm}{!}{\includegraphics{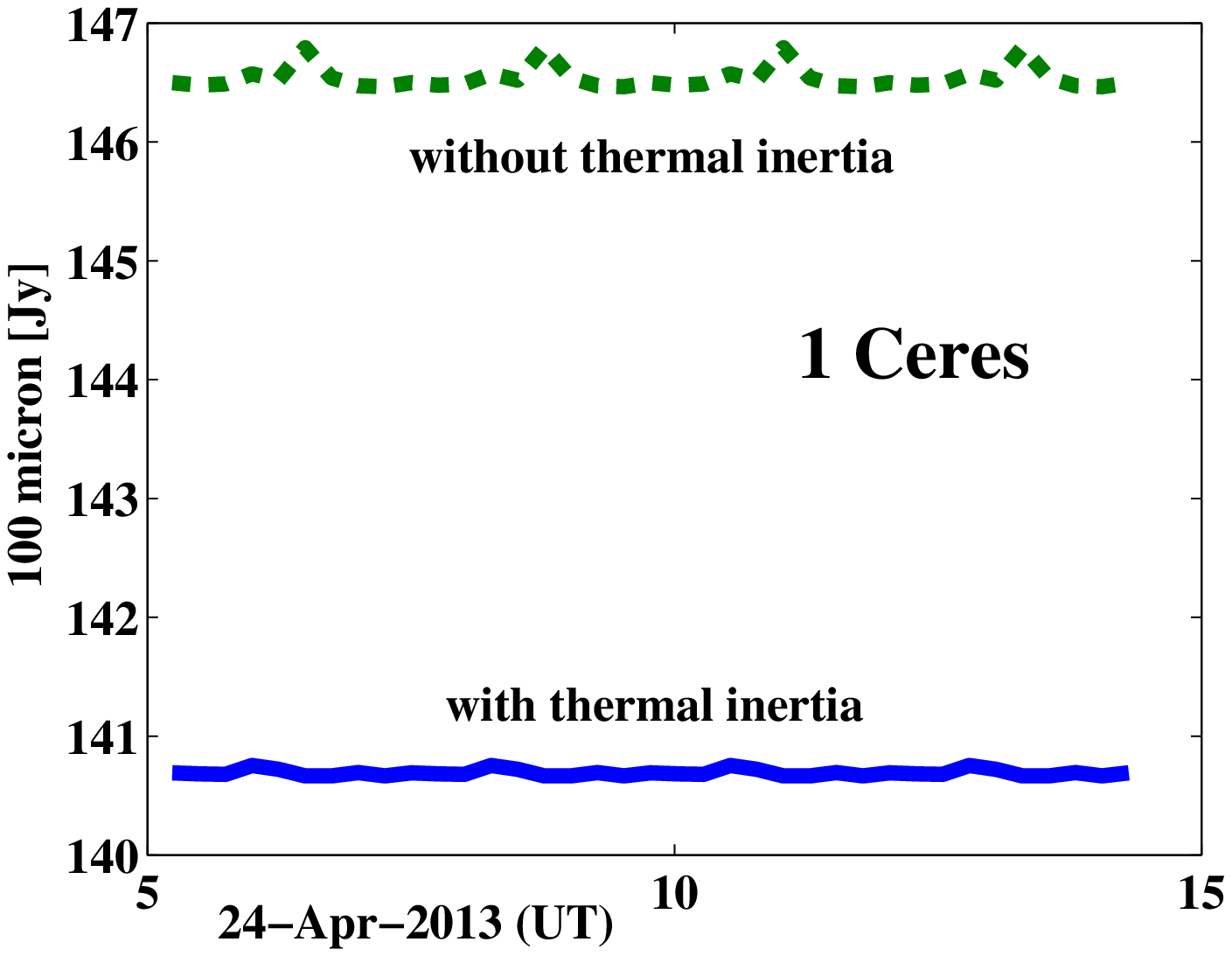}}}
  \caption{Left: Shape model of Ceres with the TPM temperature coding on the
           surface, calculated for the Herschel point-of-view on 
           OD~1441, OBSID~1342270856, rotation axis is along the
           vertical direction. Right: the corresponding thermal light-curve
           at 100\,$\mu$m with and without thermal effects included.
\label{fig:ceres_tpm}}
\end{figure}

Ceres is the largest and most massive asteroid in the main-belt. It is presumed
to be homogeneous, gravitationally relaxed and it has a low density, low albedo
and relatively featureless visible reflectance spectrum \cite{thomas05}.
The shape that best reproduces the available data (occultations, HST measurements,
adaptive optics studies, light-curve, ...) is an oblate spheroid
\cite{millis87,thomas05,carry08,drummond13} with an equatorial diameter of 974.6\,km
and a polar diameter of 909.4\,km, resulting in an equivalent diameter of
an equal volume sphere of 952.4 $\pm$ 3.4\,km \cite{thomas05}. The semi-major axes ratios
are therefore a/b = 1.0 and b/c = 1.072. The spin-axis is within 3$^{\circ}$ of
($\lambda_{sv}^{ecl}$, $\beta_{sv}^{ecl}$) = (346$^{\circ}$, +82$^{\circ}$)
\cite{drummond13} in ecliptic reference frame, with a siderial rotation period 
of 9.074170 $\pm$ 0.000001\,h \cite{chamberlain07}. The spin-vector is therefore
oriented close to perpendicular to the line-of-sight and the optical light-curve amplitude
is generally small (up to 0.04\,mag \cite{mueller98}). The geometric V-band
albedo is p$_V$ = 0.090 $\pm$ 0.0055 \cite{li06,drummond13}.
For the calculation of the Bond albedo we used an absolute magnitude H$_{V}$ = 3.28\,mag
and a slope parameter G = 0.05 \cite{lagerkvist92,mueller98}.


\clearpage
\subsection{(2)~Pallas}

\begin{figure}[h!tb]
\centering
  \rotatebox{0}{\resizebox{5.5cm}{!}{\includegraphics{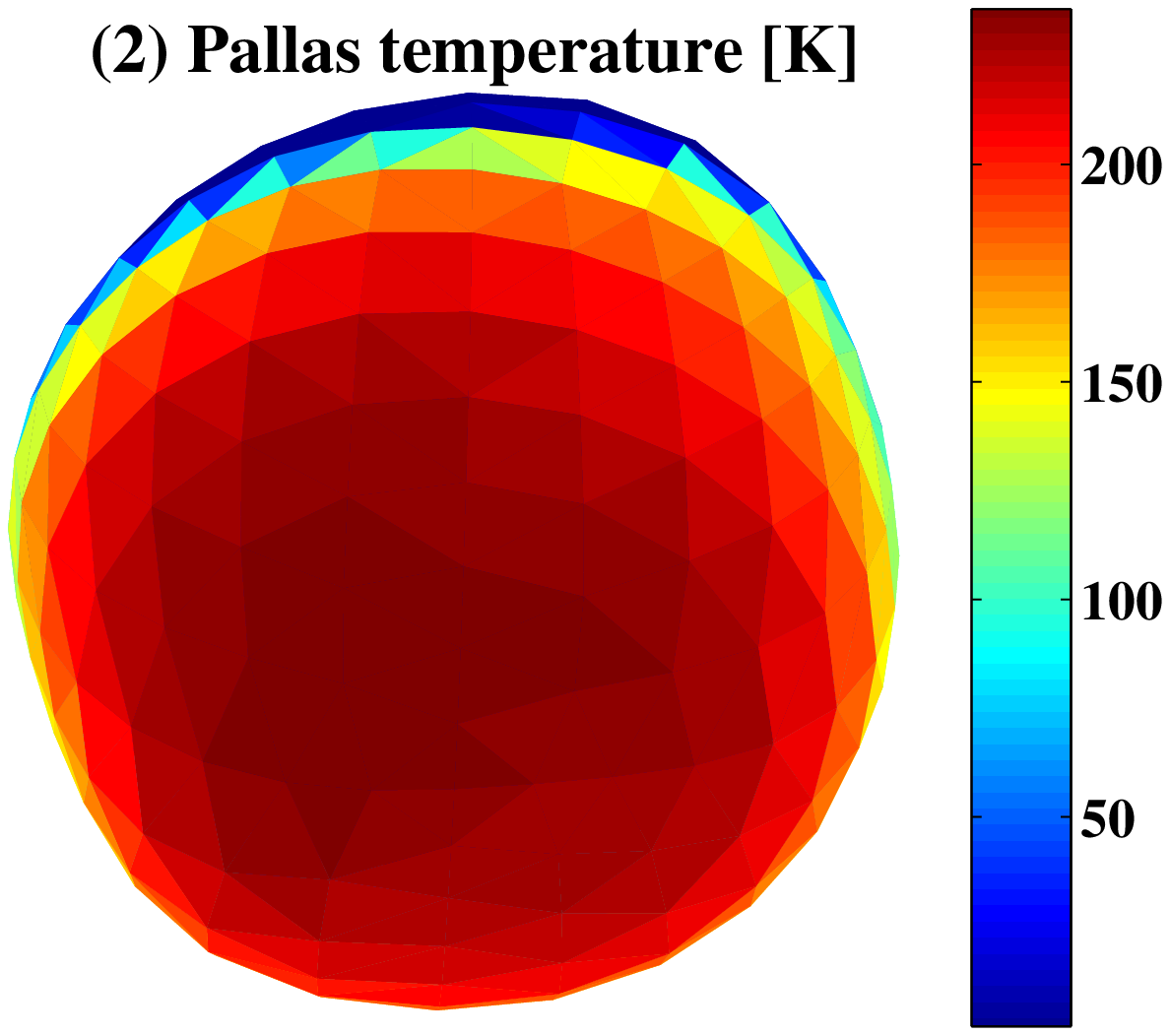}}}
  \rotatebox{0}{\resizebox{5.5cm}{!}{\includegraphics{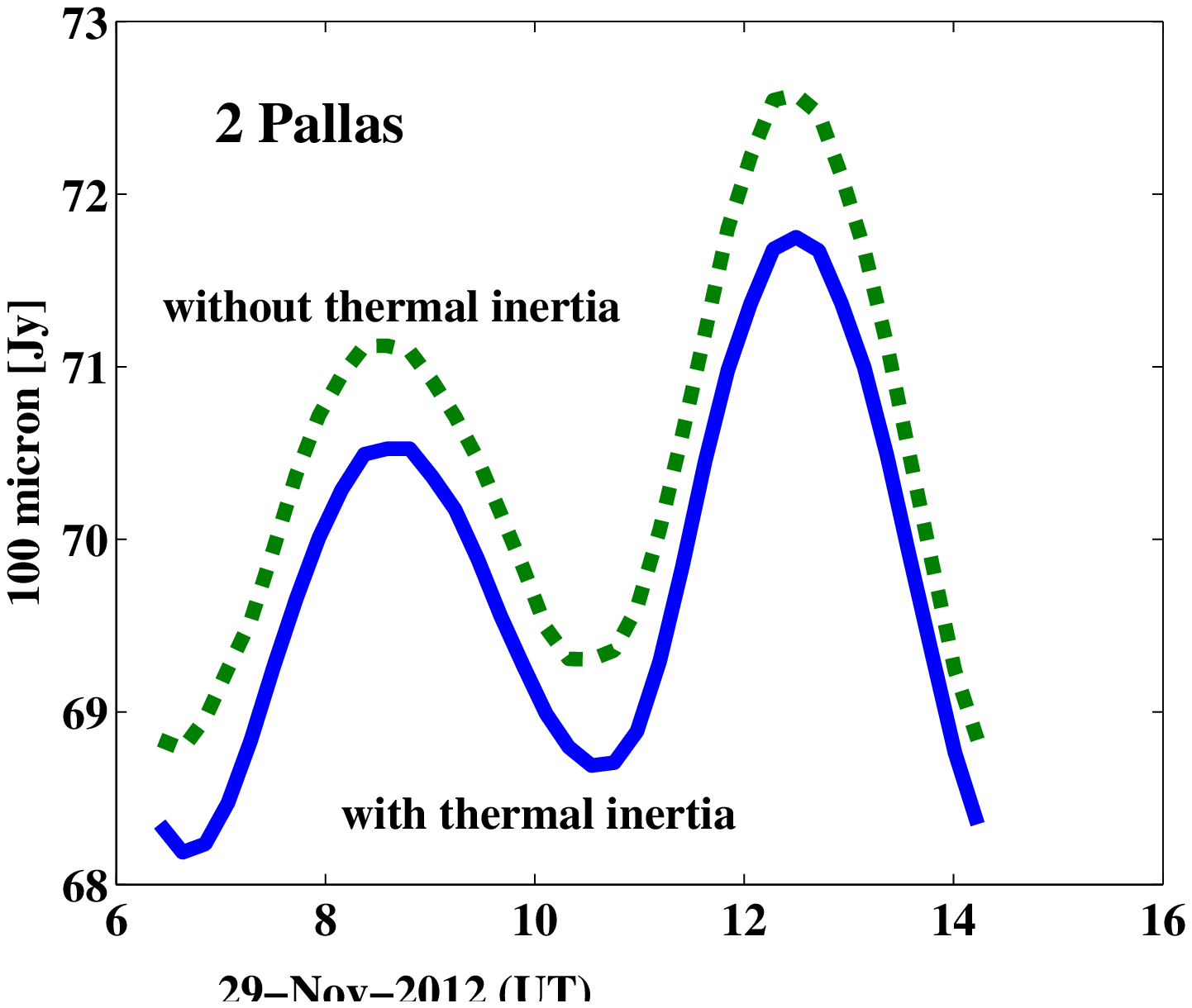}}}
  \caption{Left: Shape model of Pallas with the TPM temperature coding on the
           surface, calculated for the Herschel point-of-view on
           OD~1295, OBSID~1342256236, rotation axis is along the
           vertical direction. Right: the corresponding thermal light-curve
           at 100\,$\mu$m with and without thermal effects included.
\label{fig:pallas_tpm}}
\end{figure}

Pallas has about half the size of Ceres and is considered as an intact
protoplanet which has undergone impact excavation \cite{schmidt09}. Its 
published size, shape, and spin-properties have substantially changed over
the last years \cite{wasserman79,dunham90,schmidt09,drummond09,carry10}.
We used the latest nonconvex shape model from DAMIT\footnote{Database of Asteroid Models from Inversion Techniques,
{\tt http://astro.troja.mff.cuni.cz/projects/asteroids3D/}} with a siderial rotation period of 7.81322\,h.
This solution includes all available information from occultations, HST, light-curves over several
decades, and adaptive optics measurements. The shape can roughly be described as a
triaxial-ellipsoid body with a/b = 1.06, b/c=1.09. Its spin-axis is oriented
towards celestial directions ($\lambda_{ecl}$, $\beta_{ecl}$) = (31$^{\circ}$ $\pm$ 5$^{\circ}$,
-16$^{\circ}$ $\pm$ 5$^{\circ}$), which means it has a high obliquity of 84$^{\circ}$,
leading to high seasonal contrasts. Shape-introduced light-curve amplitudes can
reach up to 0.16\,mag \cite{apc5}. The effective size 2$\times$(abc)$^{1/3}$,
a critical parameter for our calculations, was given as 533 $\pm$ 6\,km \cite{dunham90},
545 $\pm$ 18\,km \cite{schmidt09}, 513 $\pm$ 7\,km \cite{carry10}. We adopted the first
value which has the smallest errorbar and which is based on multiple occultations,
including one of the best observed occultation of a star ever. We use
H$_{V}$ = 4.13\,mag and G = 0.16 \cite{lagerkvist92,mueller98,apc5}. Our geometric
albedo p$_V$ = 0.139 was calculated from H$_{V}$ and the effective size of 533\,km.

\clearpage
\subsection{(4)~Vesta}

\begin{figure}[h!tb]
\centering
  \rotatebox{0}{\resizebox{5.5cm}{!}{\includegraphics{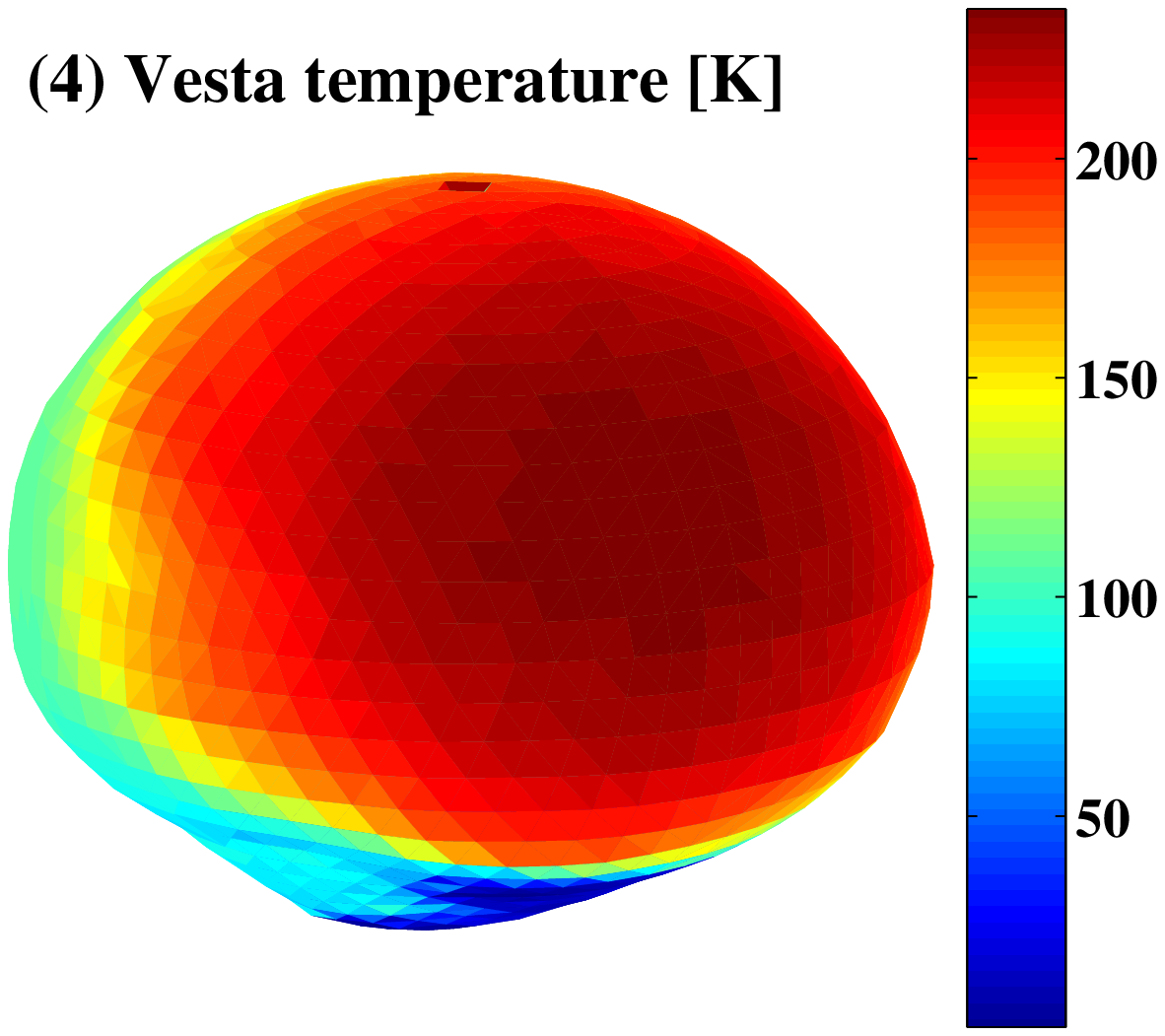}}}
  \rotatebox{0}{\resizebox{5.5cm}{!}{\includegraphics{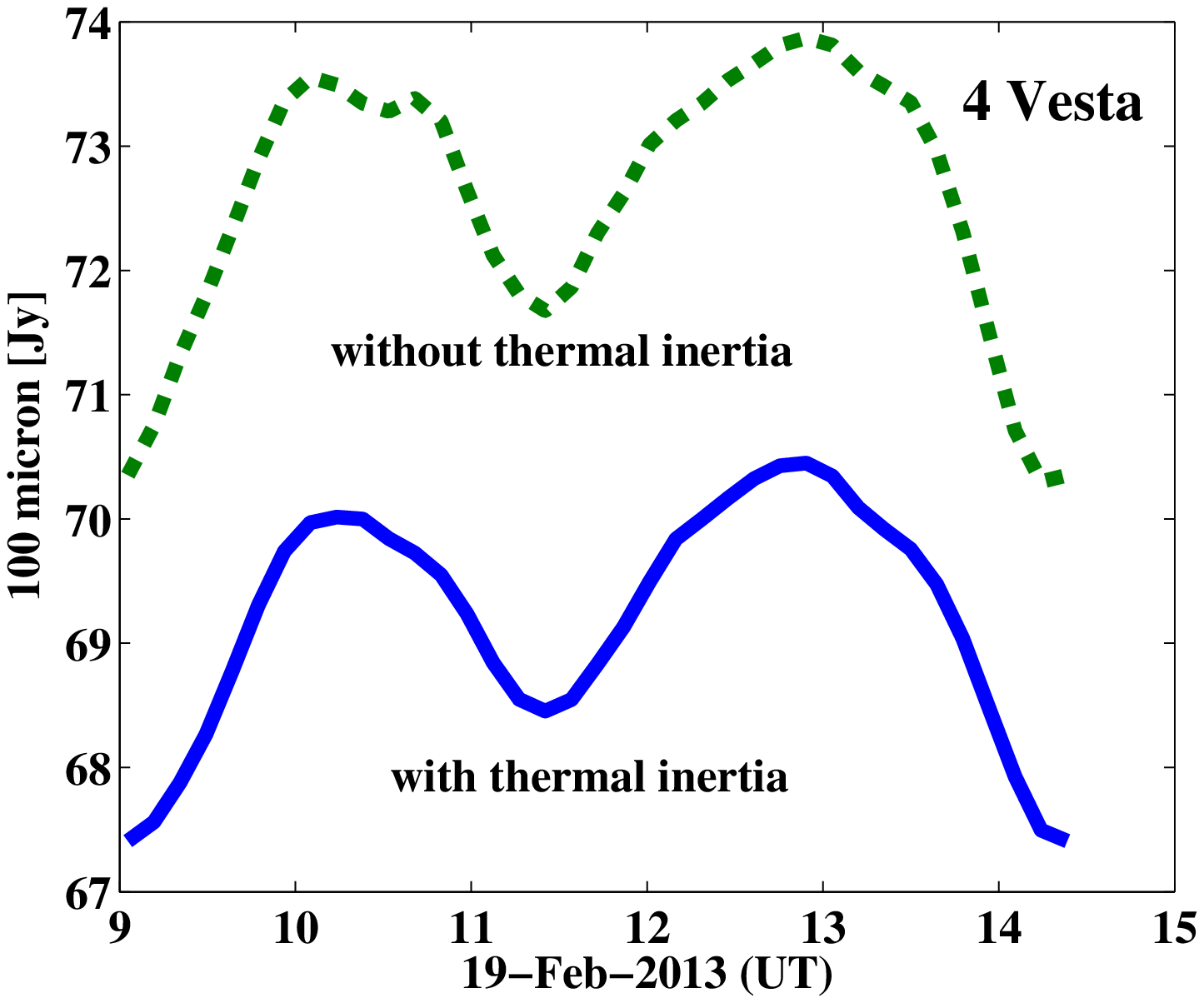}}}
  \caption{Left: Shape model of Vesta with the TPM temperature coding on the
           surface, calculated for the Herschel point-of-view on
           OD~1377, OBSID~1342263924, rotation axis is along the
           vertical direction. Right: the corresponding thermal light-curve
           at 100\,$\mu$m with and without thermal effects included.
\label{fig:vesta_tpm}}
\end{figure}

Vesta is one of the largest and the second most massive asteroid 
in the main-belt. It has recently been visited by the DAWN\footnote{\tt http://dawn.jpl.nasa.gov/}
mission. Most of the key elements for our thermophysical model purposes are
very well known, but the final shape models are not yet publically released.
Our calculations are based on the HST shape model \cite{thomas97} with a spin-vector
($\lambda_{ecl}$, $\beta_{ecl}$) = (319$^{\circ}$ $\pm$ 5$^{\circ}$,
59$^{\circ}$ $\pm$ 5$^{\circ}$), very close to values derived recently from DAWN \cite{russell12}.
The siderial rotation period is P$_{sid}$ = 5.3421289\,h \cite{thomas97,drummond88}.
The obliquity of about 27$^{\circ}$ combined with a more extreme triaxial body
leads to shape-introduced light-curve amplitudes of up to 0.18\,mag \cite{apc5}.
We assigned a mean size of 525.4 $\pm$ 0.2\,km \cite{russell12}, roughly corresponding
to a triaxial-ellipsoid body with a/b = 1.03, b/c=1.25.
We took H$_{V}$ = 3.20\,mag and G = 0.34 \cite{mueller98}, the corresponding
geometric albedo p$_V$ = 0.336 was calculated from the effective size of 525.4\,km.

\clearpage
\subsection{(21)~Lutetia}

\begin{figure}[h!tb]
\centering
  \rotatebox{0}{\resizebox{5.5cm}{!}{\includegraphics{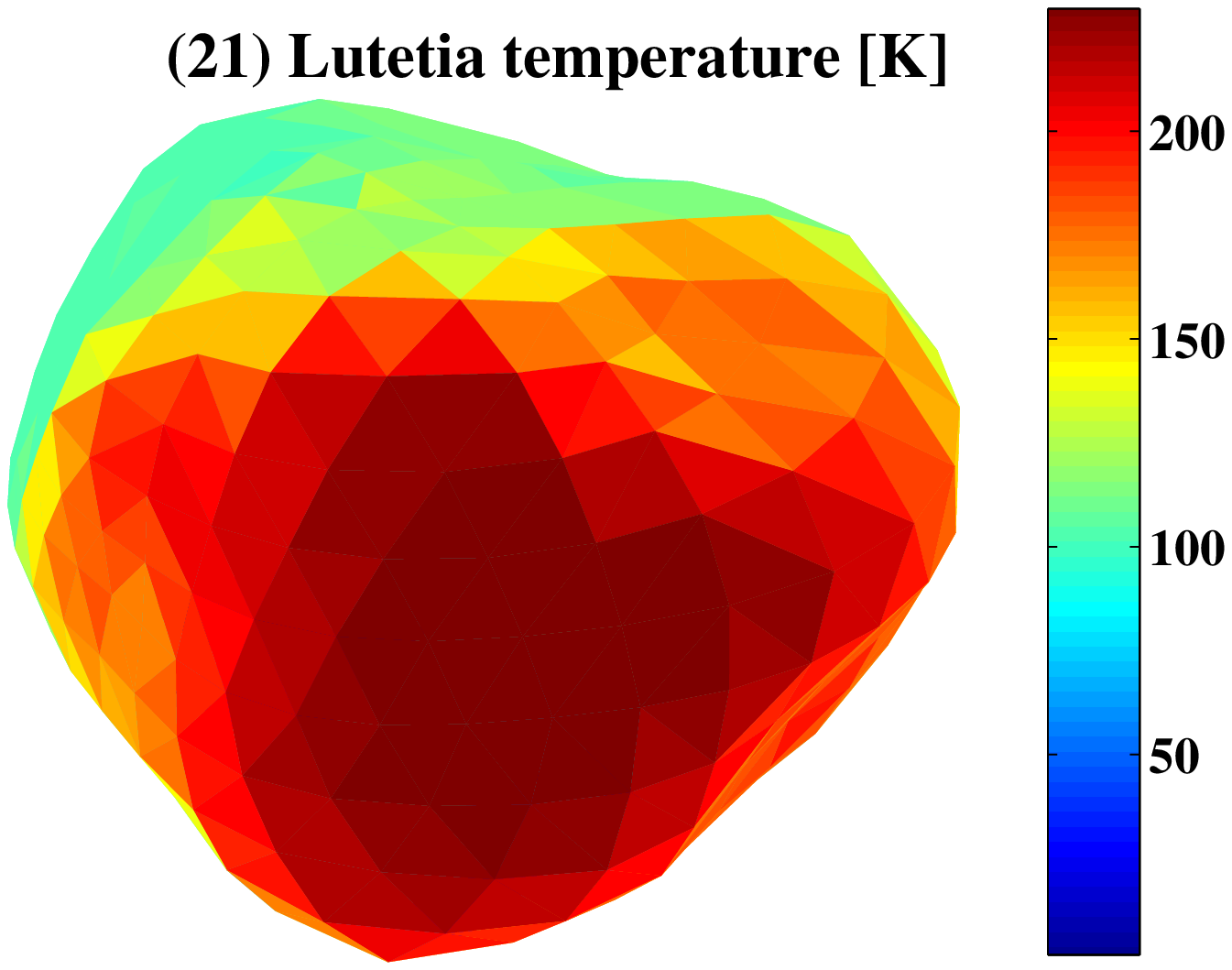}}}
  \rotatebox{0}{\resizebox{5.5cm}{!}{\includegraphics{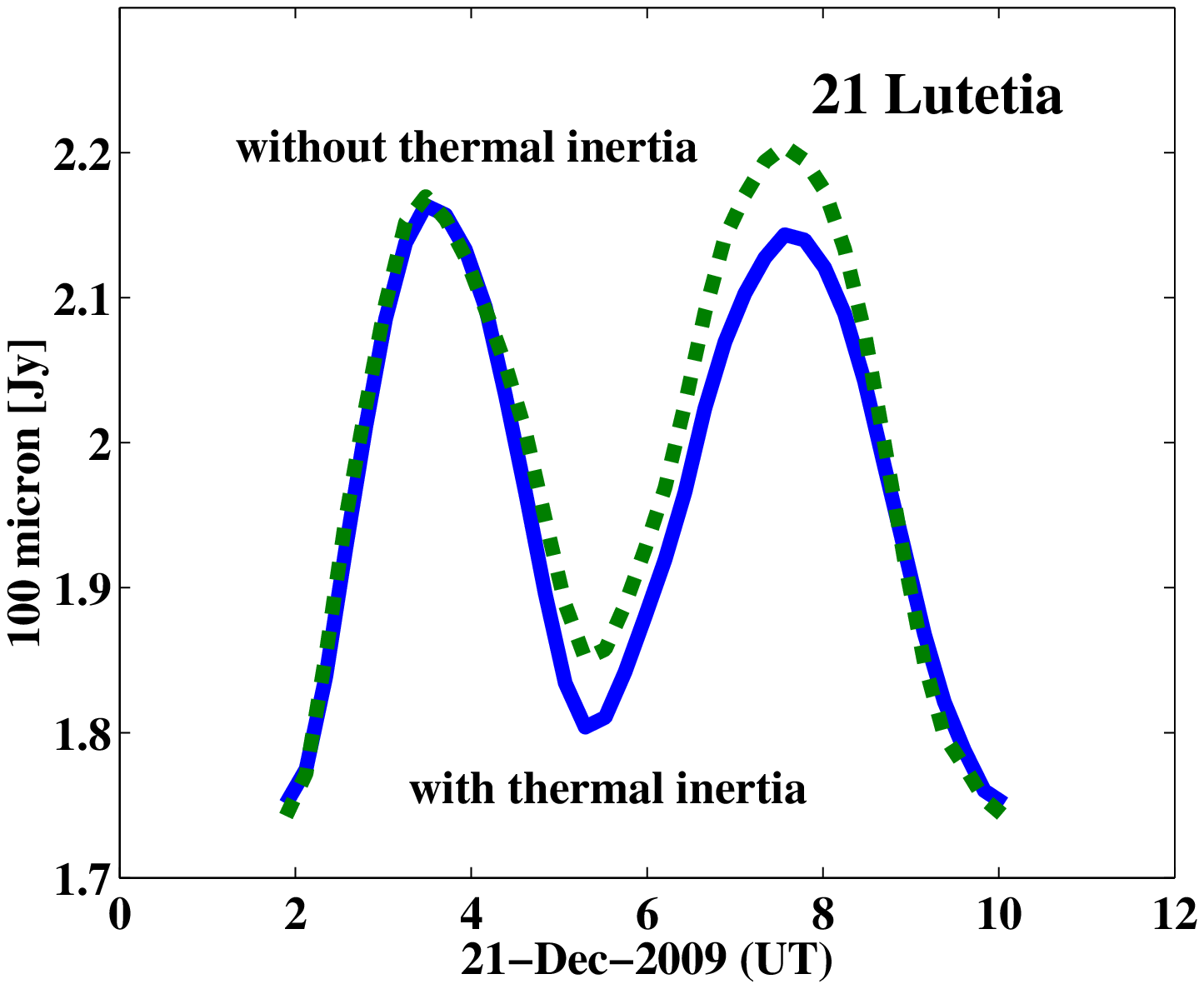}}}
  \caption{Left: Shape model of Lutetia with the TPM temperature coding on the
           surface, calculated for the Herschel point-of-view on
           OD~221, OBSID~1342188334, rotation axis is along the
           vertical direction. Right: the corresponding thermal light-curve
           at 100\,$\mu$m with and without thermal effects included.
\label{fig:lutetia_tpm}}
\end{figure}

Lutetia is significantly smaller and more irregularly shaped than the other three
objects. Due to its unusual spectral type with indications of a high metal content,
it was originally not considered in our list of potential flux calibrators. But Lutetia
was very well characterized by a ROSETTA\footnote{\tt http://www.esa.int/Our\_Activities/Space\_Science/Rosetta} flyby
in 2011 and we took advantage of the derived, high-quality properties.
Our shape model is the latest nonconvex shape model from DAMIT\footnote{Database of
Asteroid Models from Inversion Techniques, {\tt http://astro.troja.mff.cuni.cz/projects/asteroids3D/}},
which is based on a combination of flyby information, occultations, radiometry, light-curve datasets,
radar echoes, interferometry, and disk-resolved imaging \cite{carry12}.
It has a spin-vector of ($\lambda_{ecl}$, $\beta_{ecl}$) =
(52$^{\circ}$ $\pm$ 2$^{\circ}$, -6$^{\circ}$ $\pm$ 2$^{\circ}$), and a siderial rotation
period of P$_{sid}$ = 8.168271\,h \cite{lamy10,carry12}. The absolute effective size of the
final shape model is D$_{eff}$ = 99.3\,km and the measured geometric albedo is p$_V$ = 0.19 $\pm$ 0.01
\cite{carry12}. Typical shape-introduced light-curve amplitudes can reach up to 0.25\,mag \cite{apc5}.
The absolute magnitude and the slope paramemeter, both normalised to the mean
light-curve value, are given as H$_V$ = 7.25 and G = 0.12 \cite{belskaya10}.

\clearpage
\section{Results, Validity and Limitations}
\label{sec:res}

Based on the thermophysical model and object setup in Section~\ref{sec:tpm},
we calculated TPM flux densities at the PACS, SPIRE, and HIFI reference wavelengths
for the mid-time of each observation (Start-time + 0.5$\times$duration of each OBSID,
Herschel-centric reference system). The calculations have been done for the true
Herschel-centric observing geometry with the asteroid
placed at the correct helio-centric and Herschel-centric distance, under the true
phase angle and spin-vector orientation.
The observed and calibrated mono-chromatic flux densities have then been divided
by the TPM predictions. The ratios are shown in the following figures 
\ref{fig:ceresres}, \ref{fig:pallasres}, \ref{fig:vestares}, \ref{fig:lutetiares}, and are
listed in Tables~\ref{tbl:ceresres}, \ref{tbl:pallasres}, \ref{tbl:vestares}, \ref{tbl:lutetiares}, 
and discussed below.

\begin{figure}[h!tb]
\centering
  \rotatebox{90}{\resizebox{!}{\hsize}{\includegraphics{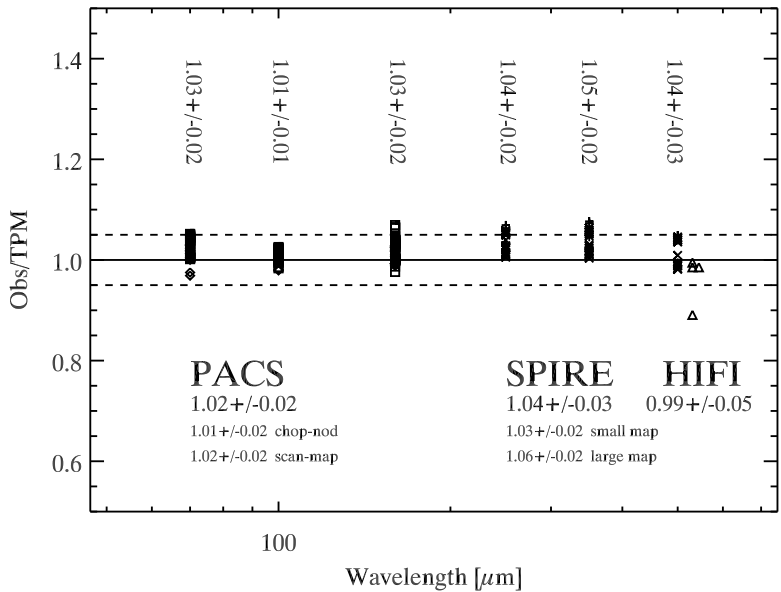}}}
  \caption{Observed and calibrated Herschel flux densities of Ceres divided by the 
           corresponding TPM predictions (one point per OBSID). The median
           ratios for each instrument and each band are given together with the
           standard deviations of the ratios. For PACS and SPIRE we also give the
           ratios per observing mode. PACS data are shown as diamonds (chop-nod data)
           and squares (scan-map data), SPIRE data are shown as plus-symbols (large map mode)
           and crosses (small map mode), HIFI data are shown as triangles.}
\label{fig:ceresres}
\end{figure}

\begin{figure}[h!tb]
\centering
  \rotatebox{90}{\resizebox{!}{\hsize}{\includegraphics{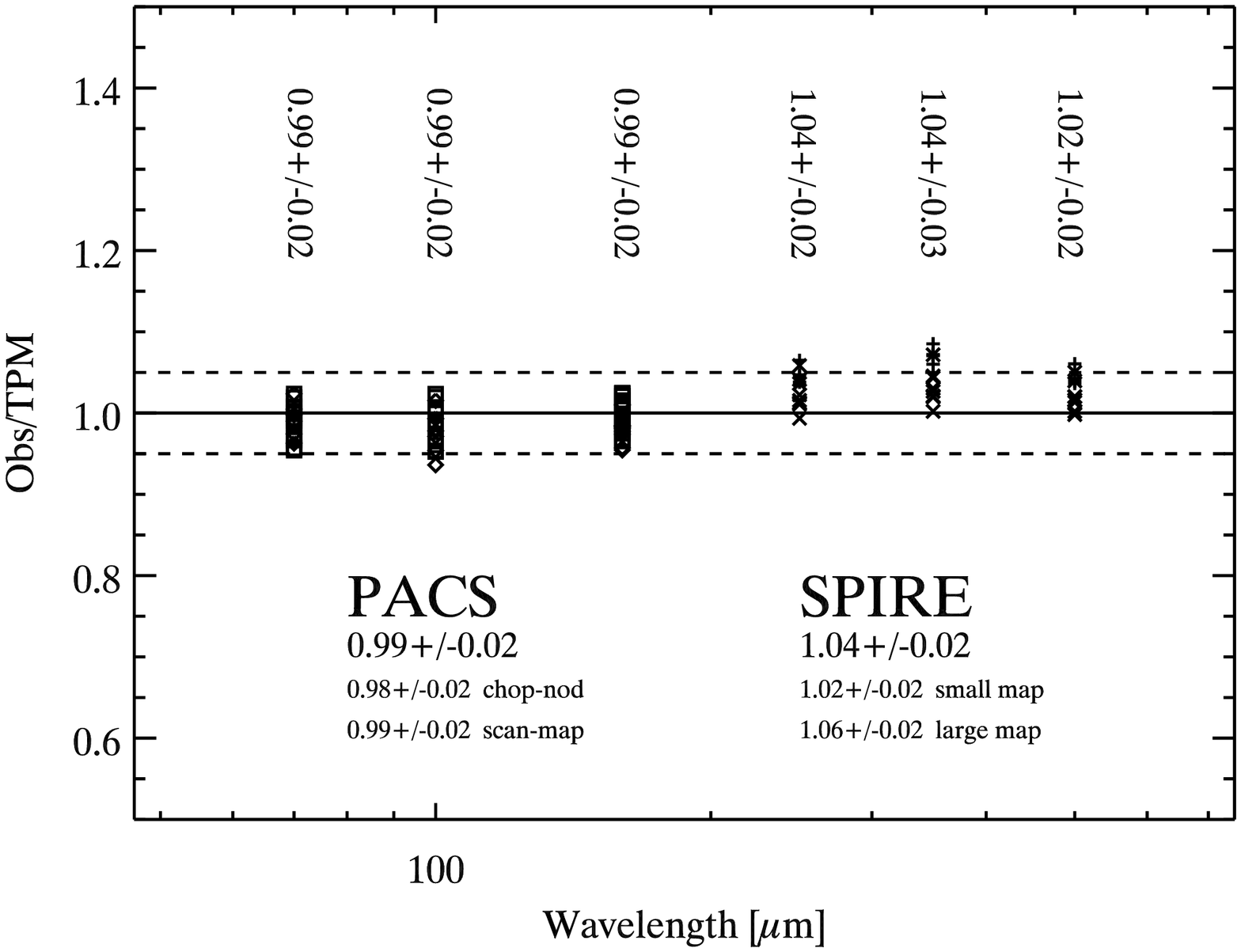}}}
  \caption{Observed and calibrated Herschel flux densities of Pallas divided by the
           corresponding TPM predictions, like in Fig.~\ref{fig:ceresres}.}
\label{fig:pallasres}
\end{figure}

\begin{figure}[h!tb]
\centering
  \rotatebox{90}{\resizebox{!}{\hsize}{\includegraphics{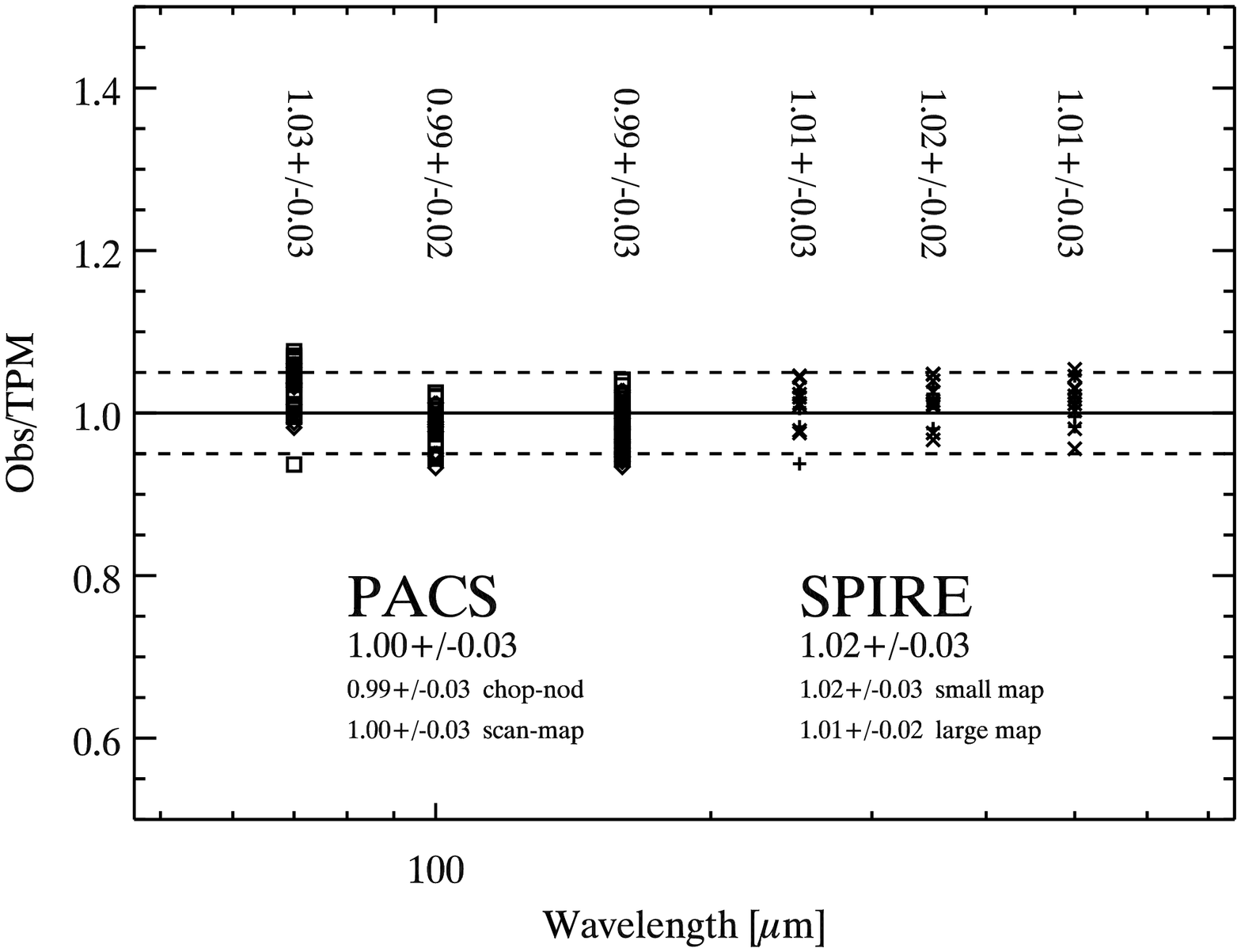}}}
  \caption{Observed and calibrated Herschel flux densities of Vesta divided by the
           corresponding TPM predictions, like in Fig.~\ref{fig:ceresres}.}
\label{fig:vestares}
\end{figure}

\begin{figure}[h!tb]
\centering
  \rotatebox{90}{\resizebox{!}{\hsize}{\includegraphics{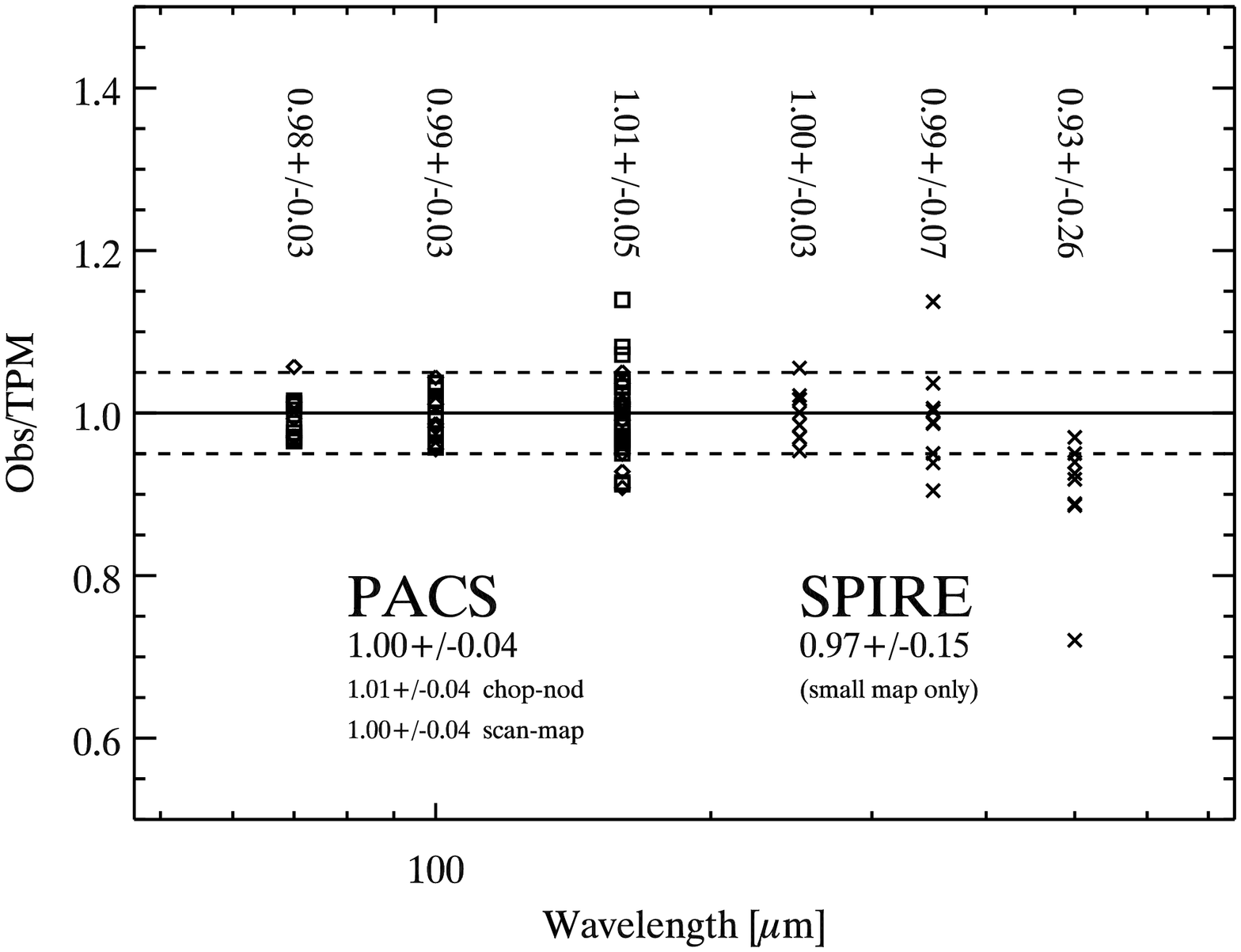}}}
  \caption{Observed and calibrated Herschel flux densities of Lutetia divided by the
           corresponding TPM predictions, like in Fig.~\ref{fig:ceresres}.}
\label{fig:lutetiares}
\end{figure}

\paragraph{Absolute flux level.}

The median ratios for all four asteroids and in all PACS \& SPIRE bands are well
within 1.00 $\pm$ 0.05. There are no systematic outliers visible. A few individual
measurements are slightly outside the 5\% boundary, but here it is not clear if
the problem is related to instrumental/technical issues or sky background related
effects. Our sources have apparent sky motions of up to about 80$^{\prime \prime}$/h
(as seen from Herschel) and they cross background sources and
dense star fields. And indeed, Ceres, Vesta and Lutetia reached galactic latitudes
below 5$^{\circ}$ during Herschel observing periods and bright sources (not easily
recognized in automatic processing) could have influenced the photometry in rare cases.
The influence of lower S/N levels can be seen in the increased ratio scatter 
in the PACS 160\,$\mu$m and SPIRE 500\,$\mu$m measurements of Lutetia.

The maximum-to-minimum observed flux ratios in a given band 
are 2.3, 2.7, 3.5, 4.3 for Ceres, Pallas, Vesta, and Lutetia, respectively
(see Tables~\ref{tbl:ceresres}, \ref{tbl:pallasres}, \ref{tbl:vestares}, \ref{tbl:lutetiares}).
This flux change is mainly dominated by changing distances between the
asteroid and Herschel, with smaller influences from changing heliocentric asteroid distances
and phase angles. The TPM setup handles these seasonal geometric
effects with high accuracy. We found no significant remaining trends in the obs/TPM-ratios
with helio-centric and Herschel-centric distance.

\paragraph{Short-term variations.}

The four asteroids are not spherical and optical lightcurves show amplitudes of up to
0.25\,mag. These variations are caused mainly by rotationally changing
cross-sections and therefore expected to be seen in disk-integrated
thermal emission as well. In our model setup this is handled by complex shape models
combined with spin-vector information (derived from occultation results,
light-curve inversion techniques, high resultion imaging techniques and/or flyby
information), and object-specific zero-points in time and rotational phase.
Our setup explains the available optical light-curves and other cross-section
related data very well, but it is not entirely clear if such models would also 
explain the rotational changes in thermal emission. Figures~\ref{fig:ceresres},
\ref{fig:pallasres}, \ref{fig:vestares}, \ref{fig:lutetiares} show -at least in
the most reliable shortest wavelength bands at 70 and 250\,$\mu$m- very small
standard deviations in the obs/TPM ratios. For Ceres and Pallas we find standard
deviations of 2\%, while for the more complex shaped objects Vesta and Lutetia
we find 3\%. This very low scatter agrees with findings on non-variable reference
stars (see \cite{balog13}) and tells us that the shape and rotational properties
of the four asteroids are modeled with sufficient accuracy.

\paragraph{Spectral shape aspects.}

The obs/TPM ratios for a given object are almost identical for all
bands of the same instrument. This is an indication that our TPM
reproduces the observed slopes in spectral energy distribution (SED)
correctly. The modeled object SEDs are summing up all different surface
temperatures over the entire disk. Here, at long wavelength and close
to the Rayleigh-Jeans SED approximation, the SEDs are closely correlated
with the disk-averaged temperatures, while at shorter wavelengths (e.g.,
in the mid-IR) the hottest sub-solar regions dominate the SED shapes.
The constant ratios over all three bands of an instrument also confirm
the validity of the strongly band-dependent colour-correction (see Sections~\ref{sec:obspacs}
and \ref{sec:obsspire}). These corrections are calculated from the 
lab-measured relative spectral response functions of the individual bands
\cite{griffin13,mueller11b}. Our results show no problems with the
tabulated colour-correction values, a nice confirmation that the bands
are well characterized and that there are no indications for filter leaks.

\paragraph{PACS/SPIRE cross-calibration and emissivity aspects.}

For the 3 bright sources Ceres, Pallas, and Vesta the SPIRE ratios are
2-5\% higher than the PACS ratios. The cause is not clear, but there are
different possibilities:
(i) A systematic difference in the absolute
flux calibrators (5 fiducial stars for PACS \cite{balog13} and a specific
Neptune model for SPIRE \cite{bendo13}). Both calibration systems are given
with an absolute accuracy of $\pm$5\% and the offset we see in the asteroids
is within this range. Both calibration systems underwent recent adjustments
and re-adjustments with typical changes of a few percent. Discussions are
still ongoing and the related publications -possibly with slight adjustments-
are in preparation.
(ii) A flux-dependency in the reduction/calibration
steps which is not correctly accounted for: the PACS asteroid data are corrected
for detector non-linearities (up to 6\% for the highest asteroid fluxes), but
an absolute validation at these flux levels is difficult. The SPIRE asteroid
data are well below the flux level of Neptune which is used as reference object
and the asteroids also move much faster than Neptune on the sky. Both aspects might cause
an offset of a few percent on the final fluxes.
(iii) The asteroid models use
a wavelength-dependent emissivity model \cite{mueller98,mueller99,mueller02}
and the largest emissivity changes happen between 200 and 500\,$\mu$m. But
if there are problems in the emissivity model solution we would expect to 
see obs/TPM ratios changing gradually with wavelengths and not in a step-function
as we see it here.
We also tested a constant emissivity model ($\epsilon$ = 0.9 = const.) which
clearly confirms that lower and wavelength-dependent emissivities are needed
to explain the SPIRE measurements.
However, the effective emissivity changes are not precisely known for the
region beyond $\approx$150\,$\mu$m where subsurface layers (probably with
different thermal properties \cite{keihm84}) start to become visible.
A future scientific analysis of all combined PACS and SPIRE
observations might reveal a new and slightly different wavelength-dependent
emissivity model for the large main-belt asteroids.

One additional element in
this context is the outcome of a dedicated PACS/SPIRE cross-calibration study
for the fiducial calibration stars and -at a much higher flux level- for the
planets Uranus and Neptune. In this way, one could investigate further the
reason for the small jump between PACS and SPIRE fluxes.\\

There are a few additional points which deserve mentioning:
\begin{itemlist}
\item[$\bullet$] For Lutetia we find significantly larger standard deviations in
                 the obs/TPM ratios at 350 and 500\,$\mu$m, but Lutetia is already
		 faint at these wavelength (below 600\,mJy at 350\,$\mu$m and some
		 measurement are even below 100\,mJy at 500\,$\mu$m) and
		 background contamination and instrument noise levels start to contribute.
\item[$\bullet$] The 70\,$\mu$m obs/TPM ratio for Vesta is about 4\% higher than
                 the ratios at 100 and 160\,$\mu$m. We don't know the reason for this
		 effect, but we speculate that this might be the result of a broadband mineralogic
		 surface feature covered by the 70\,$\mu$m-band ($\sim$55 - 95\,$\mu$m). We plan
		 to follow this up via PACS spectrometer measurements of Vesta.
\item[$\bullet$] For the 3 brightest targets we also see a very small difference between
                 PACS data taken in chop-nod mode and scan-map mode. The chop-nod ratios are
		 about 1\% lower than the corresponding scan-map ratios. We expected to
		 see slightly underestimated fluxes in the chop-nod mode for very bright
		 targets (see \cite{nielbock13}), but it was not clear how big the effect
		 would be. Based on the asteroid results, we expect to see a 2-3\% flux
		 differences for even brighter targets, like Neptune and Uranus, between
		 measurements taken in these two different PACS observing modes.
\item[$\bullet$] The HIFI ratios are very close to the PACS ratios and about 5\% lower than
                 the SPIRE ratios. But the derived fluxes are very sensitive to pointing
		 errors. Additional pointing errors could increase the derived
                 flux densities by up to $\approx 5$\% which would then bring the
		 HIFI ratios very close to the SPIRE ones.
\item[$\bullet$] There was one SPIRE observation of Vesta (OD 411, OBSID 1342199329, Large Map mode)
                 which produced fluxes which are about 35\% higher than the corresponding
                 model predictions in all 3 bands, probably due to a contaminating background
                 source. We eliminated this measurement from our analysis.
\item[$\bullet$] We see a 3-4\% offset between SPIRE large and small scan map observations
                 of Ceres and Pallas, but not for Vesta. This offset is not present in
                 observations taken in both modes close in time. The cause is therefore
                 either related to satellite/instrument effects changing with time 
                 (like the changing telescope flux) or by a time-dependent thermal effect
                 which is not covered by our current model-setup. A first investigation
                 seems to point towards a small effect related to subsurface emission which
                 seems to play a role at the longest SPIRE wavelengths and which is at
                 present only approximated by our wavelength-dependent emissivity models.
\item[$\bullet$] Ceres, Pallas, and Vesta also have SPIRE observations taken in non-standard
                 scanning mode. A first comparison with our model predictions confirms
		 the validity of the data. They will be included in future analysis projects.
\end{itemlist}

\paragraph{Quality of model parameters.}

The fact that our TPM predictions agree -on absolute scale- very well with the
Herschel measurements does not automatically mean that all our object properties
(mainly effective size, albedo, thermal properties) are correct. The object-related
quantities have uncertainties and could be even slightly off. But we aimed for finding
the most accurate object sizes, and derived preferentially from direct measurements
and published in literature.
The thermal properties influence the predictions in an absolute sense and also
in a wavelength-dependent manner. We took default values from literature to avoid
any dependency of our object properties from Herschel-related information.
Overall, our model settings allow us to reproduce the observed absolute fluxes
and SED shapes with high accuracy and we have therefore great confidence in our
model solutions.

\paragraph{Accuracy.}

 \begin{figure}[h!tb]
\centering
  \rotatebox{0}{\resizebox{\hsize}{!}{\includegraphics{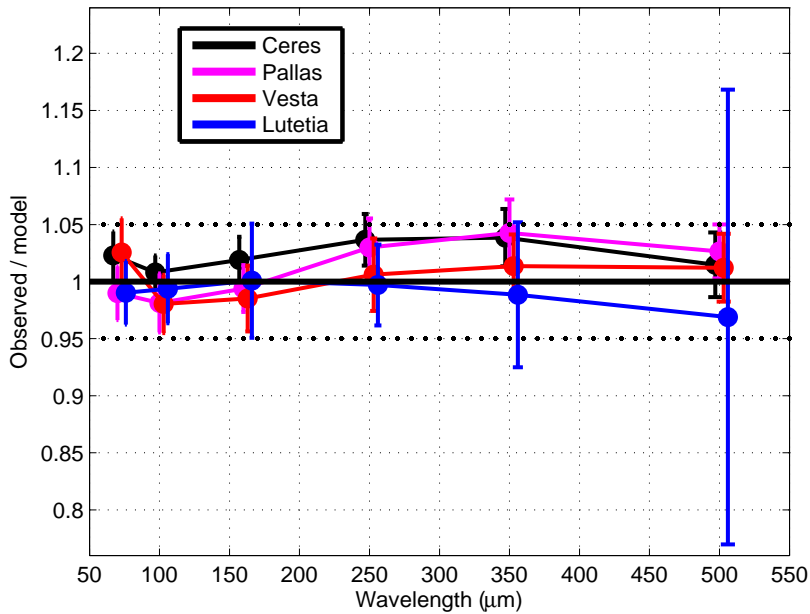}}}
  \caption{Dispersion in the ratios of measured-to-model fluxes for the four
           asteroids as a function of wavelength. The weighted mean ratios are
           shown with errorbars reflecting the absolute flux calibration of
           individual measurements as well as the variance of the sample.}
\label{fig:ratios}
\end{figure}

We made a statistical analysis of the observation-to-model ratios for all four asteroids
(Table~\ref{tbl:ratios}, Figure~\ref{fig:ratios}) to see how many observational data points are matched by the
corresponding TPM prediction. 

\begin{table}
\begin{center}
\caption{Statistical comparison between observed and TPM fluxes. The numbers
         indicate how many observations are matched by the corresponding
         TPM prediction within the given 1-$\sigma$ error bars and how many are
         not matched. The last two lines give the agreement per band in percent,
         based on the 1-$\sigma$ and 2-$\sigma$ errors in the observed fluxes.}
\label{tbl:ratios}
\begin{tabular}{rlrrrlrrr}
\hline\noalign{\smallskip}
\multicolumn{2}{c}{Object} &  70\,$\mu$m & 100\,$\mu$m  &  160\,$\mu$m &  250\,$\mu$m &  350\,$\mu$m &  500\,$\mu$m\\
\noalign{\smallskip}\hline\noalign{\smallskip}
 1 & Ceres   & 51/0  &      39/0  &      84/4  &     20/4  &   18/6   &     24/0 \\
 2 & Pallas  & 20/0  &      19/1  &      40/0  &     11/2  &    8/5   &     12/1 \\
 4 & Vesta   & 32/8  &      25/7  &      64/8  &     14/1  &   15/0   &     15/0 \\
21 & Lutetia & 14/1  &      19/0  &      29/5  &      9/0  &    6/3   &      5/4 \\
\noalign{\smallskip}\hline\noalign{\smallskip}
\multicolumn{2}{c}{1-$\sigma$ agreement} &  92\% & 92\% &  92\% & 87\% &  70\%  & 91\% \\
\multicolumn{2}{c}{2-$\sigma$ agreement} & 100\% & 100\% &  100\% & 100\% &  100\%  & 97\% \\
\noalign{\smallskip}\hline
\end{tabular}
\end{center}
\end{table}

For this comparison we considered the absolute flux errors for each
individual measurement as it was produced by the general data reduction
and calibration procedure mentioned above. The absolute flux errors include
the processing errors, the photometry errors (corrected for correlated noise)
and the absolute flux calibration error as provided by the
three instrument teams. In all three PACS bands more than 90\% of all measurements
agree within their 1-$\sigma$ errorbars with the corresponding TPM prediction. In the
SPIRE bands the agreement is still between about 70\% and 90\%. If we allow for
2-$\sigma$ errorbars we find 100\% agreement for Ceres, Pallas and Vesta in all
six bands. For Lutetia there are only two measurements in the 500\,$\mu$m
band which are outside the 2-$\sigma$ threshold.
Figure~\ref{fig:ratios} shows the agreement between observations and TPM predictions
in a graphical way. For each object we calculated the weighted mean ratio and errorbars
adding up quadratically the variance of the weighted mean and the weighted sample
variance. These errorbars are dominated by the 5\% absolute flux calibration errors
of our measurements, the variance of the weighted mean (as can be seen in 
Figures~\ref{fig:ceresres}, \ref{fig:pallasres}, \ref{fig:vestares}, \ref{fig:lutetiares})
is typically 2-3\% only, with exception of the long-wavelength channels for Lutetia.
This excellent agreement between observed and TPM fluxes confirms the validity of the
four asteroids as prime calibrators on a similar quality level as given for the fiducial star
models in the PACS range (5\%), the Neptune model in the SPIRE range (5\%),
and the Mars model in the HIFI band 1a/1b (5\%).

\paragraph{Limitations.}

Our comparison between TPM predictions and measurements is limited to a
wavelength range between about 50\,$\mu$m (short wavelength end of the PACS 70\,$\mu$m
filter) and about 700\,$\mu$m (long wavelength end of the SPIRE 500\,$\mu$m
filter). The Herschel visibility constrained the tested phase angles to values
between about 15$^{\circ}$ and 30$^{\circ}$ before and after opposition. Outside
these wavelengths and phase angle ranges the TPM might have slightly higher
uncertainties. Some of the asteroids have complex shapes and the shape models
used might not characterize the true shape very accurately.
This could also cause small deviations between TPM predictions and
the true measured fluxes for specific viewing geometries. The rotation periods
are known with high accuracy for all four asteroids and the applied spin vectors are
of sufficient quality for the next decade.

Overall, the TPM deviations outside the specified wavelengths and phase-angle
ranges are expected to be small and absolute model accuracies of better than
10\% seem to be reasonable for all four asteroids. Further testing against
additional thermal data is foreseen in the near future to cover the full
ALMA and SPICA\footnote{\tt http://www.ir.isas.jaxa.jp/SPICA/SPICA\_HP/index\_English.html} regime.

\section{Conclusions}
\label{sec:con}

We find the following general results related to the 4 asteroids:
\begin{itemlist}
\item[$\bullet$] The new asteroid models predict the observed fluxes on absolute scales
                 with better than 5\% accuracy in the in the 50 to 700\,$\mu$m range. This
                 means that the effective size and albedo values in our model setup are
                 of high quality.
\item[$\bullet$] Shape and spin properties dominate the short-term brightness variations:
                 our shape, rotation-period and spin-axis approximations are sufficient
                 for our purposes.
\item[$\bullet$] In general, the rotational and seasonal flux changes are modeled with
                 high quality to account for short-term (rotational effects on time scales of
                 hours) and long-term (effects with phase angle and changing distance to the Sun on
                 time scales of months or years) object variability.
\item[$\bullet$] Our ``default" description of the thermal properties is sufficient to
                 explain the observed far-IR/sub-mm fluxes. Please note that the Vesta
                 emissivity is very different from the emissivity model used for the
                 other objects.
\item[$\bullet$] The asteroid surfaces of all 4 asteroids are very well described by a 
                 low-conductivity, hence low thermal inertia surface regolith with very
                 little heat transport to the nightside of the object (they are observed
                 at phase angles up to about 30$^{\circ}$).
\item[$\bullet$] There are indications that Vesta has a broad shallow mineralogic emission
                 feature which contributes up to 4\% to the total flux measured in the
		 PACS 70\,$\mu$m band.
\end{itemlist}

We find the following Herschel-related results:
\begin{itemlist}
\item[$\bullet$] The PACS chop-nod and scan-map derived fluxes agree very well (within 1\%),
                 although there seems to be a small tendency that the chop-nod fluxes
                 are slightly underestimated for very bright targets.
\item[$\bullet$] The SPIRE observation/model ratios for Ceres, Pallas, and Vesta are
                 2-5\% higher than the PACS related ones. This could be related to
                 the very different calibration schemes of both instruments, but there
                 is also the possibility of a model-introduced effect (e.g., related
                 to the object emissivity models).
\item[$\bullet$] The reduced and calibrated HIFI continuum fluxes for Ceres agree
                 very well with the PACS measurements and confirm the high photometric
		 quality of the HIFI continuum measurements.
\end{itemlist}

Ceres, Pallas, Vesta, and Lutetia as prime calibrators:
\begin{itemlist}
\item[$\bullet$] The new TPM setup for the four asteroids predict the observed fluxes
                 on absolute scales with better than 5\% accuracy in the
                 wavelength range 50 to 700\,$\mu$m and for phase angles between
                 $\sim$15$^{\circ}$ and $\sim$30$^{\circ}$. 
\item[$\bullet$] Outside the Herschel PACS/SPIRE wavelength range and for extreme phase
                 angles we still expect that the absolute accuracy of the TPM predictions are better
                 than 10\%.
\end{itemlist}

Overall, our thermophysical model predictions for the four asteroids agree within
5\% with the available (and independently calibrated) Herschel measurements.
The achieved absolute accuracy is similar to the ones quoted for the
official Herschel prime calibrators, the stellar photosphere models, the Neptune and Mars
planet models, which justifies to upgrade the four asteroid models to the rank of
prime calibrators.
The four objects cover the flux regime from just below 1,000\,Jy (Ceres at mid-IR) down
to fluxes below 0.1\,Jy (Lutetia at the longest wavelengths). Based on the comparison with
PACS, SPIRE and HIFI measurements and pre-Herschel experience, the validity
of prime calibrators ranges from mid-infrared to about 600\,$\mu$m, connecting nicely
the absolute stellar reference system in the mid-IR with the planet-based calibration
at sub-mm/mm wavelengths.

\begin{acknowledgements}
We would like to thank the PIs of the various scientific projects for permission
to use their Herschel science data in the context of our calibration work.
\end{acknowledgements}


\clearpage
\appendix
\section*{Overview of available Herschel photometric measurements}
\label{sec:obsapp}

In the following tables we list the available photometric observations (calibration and science observations)
with one of the four asteroids in the field of view. Some of the early measurements were used with
very different instrument settings and non-standard observing modes. The corresponding fluxes are not
well calibrated and we excluded them from our analysis.\\

\label{tbl:ceresobs1}
{\it Table~2 "Overview of all relevant Herschel-PACS photometer scan-map observations of (1)~Ceres" is available in the Exp.\ Astron.\ online version.}\\

\label{tbl:ceresobs2}
{\it Table~3 "Overview of all relevant Herschel-PACS photometer chop-nod observations of (1)~Ceres" is available in the Exp.\ Astron.\ online version.}\\

\label{tbl:ceresobs3}
{\it Table~4 "Overview of all relevant Herschel-SPIRE photometer observations of (1)~Ceres" is available in the Exp.\ Astron.\ online version.}\\

\label{tbl:ceresobs4}
{\it Table~5 "Overview of all relevant Herschel-HIFI point observations of (1)~Ceres" is available in the Exp.\ Astron.\ online version.}\\

\label{tbl:pallasobs1}
{\it Table~6 "Overview of all relevant Herschel-PACS photometer scan-map observations of (2)~Pallas" is available in the Exp.\ Astron.\ online version.}\\

\label{tbl:pallasobs2}
{\it Table~7 "Overview of all relevant Herschel-PACS photometer chop-nod observations of (2)~Pallas" is available in the Exp.\ Astron.\ online version.}\\

\label{tbl:pallasobs3}
{\it Table~8 "Overview of all relevant Herschel-SPIRE photometer observations of (2)~Pallas" is available in the Exp.\ Astron.\ online version.}\\

\label{tbl:vestaobs1}
{\it Table~9 "Overview of all relevant Herschel-PACS photometer scan-map observations of (4)~Vesta" is available in the Exp.\ Astron.\ online version.}\\

\label{tbl:vestaobs2}
{\it Table~10 "Overview of all relevant Herschel-PACS photometer chop-nod observations of (4)~Vesta" is available in the Exp.\ Astron.\ online version.}\\

\label{tbl:vestaobs3}
{\it Table~11 "Overview of all relevant Herschel-Spire photometer observations of (4)~Vesta" is available in the Exp.\ Astron.\ online version.}\\

\label{tbl:lutetiaobs1} 						
{\it Table~12 "Overview of all relevant Herschel-PACS photometer scan-map observations of (21)~Lutetia" is available in the Exp.\ Astron.\ online version.}\\

\label{tbl:lutetiaobs2}
{\it Table~13 "Overview of all relevant Herschel-PACS photometer chop-nod observations of (21)~Lutetia" is available in the Exp.\ Astron.\ online version.}\\

\label{tbl:lutetiaobs3}
{\it Table~14 "Overview of all relevant Herschel-SPIRE photometer observations of (21)~Lutetia" is available in the Exp.\ Astron.\ online version.}\\

\label{tbl:fixedobs}
{\it Table~15 "Additional Herschel fixed position photometer observations (no tracking)" is available in the Exp.\ Astron.\ online version.}\\

\clearpage
\section*{Observational results of the Herschel photometric measurements}
\label{sec:resapp}

In the following tables we list all extracted photometric fluxes, calibrated against stars (PACS), Neptune (SPIRE),
and Mars (HIFI).\\

\label{tbl:ceresres}
{\it Table~16 "Photometric Herschel data of (1)~Ceres" is available in the Exp.\ Astron.\ online version.}\\

\label{tbl:pallasres}
{\it Table~17 "Photometric Herschel data of (2)~Pallas" is available in the Exp.\ Astron.\ online version.}\\

\label{tbl:vestares}
{\it Table~18 "Photometric Herschel data of (4)~Vesta" is available in the Exp.\ Astron.\ online version.}\\

\label{tbl:lutetiares}
{\it Table~19 "Photometric Herschel data of (21)~Lutetia" is available in the Exp.\ Astron.\ online version.}\\

\end{document}